%% file: publication.tex
\documentclass[final,leqno,onefignum,onetabnum]{siamltex1213}

\usepackage[utf8x]{inputenc}
\usepackage[english]{babel}

\usepackage{blindtext}

\usepackage{amsmath,amssymb}
\usepackage{bm}

\usepackage{graphicx}
\usepackage[caption=false]{subfig}

\usepackage{environ}
\usepackage{booktabs}
\usepackage{multirow}

\usepackage{cite}

\begin{document}


\input{front.tex}


\input{intro.tex}

\input{theory.tex}

\input{results.tex}

\input{discussion.tex}


\input{acknowledge.tex}



\bibliographystyle{siam}
\bibliography{publication}{}

\end{document}

%% file: front.tex
\title{Reversible Markov chain estimation using convex-concave programming}

\author{
  Benjamin Trendelkamp-Schroer \footnotemark[2]\ \footnotemark[3] \and
  Hao Wu \footnotemark[2]\ \footnotemark[4] \and
  Frank Noe \footnotemark[2]\ \footnotemark[5]
}




\maketitle

\renewcommand{\thefootnote}{\fnsymbol{footnote}}

\footnotetext[2]{Institut f\"{u}r Mathematik und Informatik, Freie Universit\"{a}t Berlin, Arnimallee 6, 14195 Berlin}
\footnotetext[3]{B. T.-S. was supported by Deutsche Forschungsgemeinschaft (DFG) Grant No. SFB 740}
\footnotetext[4]{H. W. was supported by DFG Grant No. SFB 1114}
\footnotetext[5]{F. N. was supported by European Research Council (ERC) starting grant pcCell}

\renewcommand{\thefootnote}{\arabic{footnote}}

\slugger{mms}{xxxx}{xx}{x}{x--x}

\begin{abstract}
  We present a convex-concave reformulation of the reversible Markov
  chain estimation problem and outline an efficient numerical scheme
  for the solution of the resulting problem based on a primal-dual
  interior point method for monotone variational
  inequalities. Extensions to situations in which information about
  the stationary vector is available can also be solved via the
  convex-concave reformulation. The method can be generalized and
  applied to the discrete transition matrix reweighting analysis
  method to perform inference from independent chains with specified
  couplings between the stationary probabilities. The proposed
  approach offers a significant speed-up compared to a fixed-point
  iteration for a number of relevant applications.
\end{abstract}




%% file: intro.tex
\section{Introduction}
The study of reversible Markov chains is a recurrent theme in
probability theory with many important applications, \cite{aldous2014,
  metropolis1953, robert2013}. Surprisingly, statistical inference for
reversible Markov chains has been studied only recently. The
reversible maximum likelihood estimation (MLE) problem was previously
discussed in \cite{bowman2009, prinz2011,
  TrendelkampWuNoe2015}. \cite{diaconis2006, noe2008,
  MetznerNoeSchuette2009, besag2013, TrendelkampWuNoe2015} study the
the posterior ensemble of reversible stochastic matrices and discuss
algorithms for Bayesian posterior inference. For a given stochastic
matrix the best approximation which is reversible with respect to a
given stationary vector was found in \cite{nielsen2015}. 

Maximum likelihood estimation and posterior inference of reversible
stochastic matrices have important applications in the context of
Markov state models \cite{bowman2013}. Markov state models are
simplified kinetic models for the complex dynamics of biomolecules.
Transition probabilities between relevant molecular conformations are
estimated from simulation data. The estimated transition matrix is
then used to compute quantities of interest and to extract a
simplified picture of the kinetic pathways present in the dynamics. In
\cite{trendelkampnoe2014} it is shown that a significant speed-up in
the estimation of rare events is possible if additional information
about the stationary vector is incorporated via a detailed balance
constraint.

The reversible MLE problem was previously solved using a self
consistent iteration method which can require a large number of
iterations to converge \cite{bowman2009, prinz2011,
  TrendelkampWuNoe2015}. Here we outline an efficient numerical
algorithm for solving the reversible MLE problem via a convex-concave
reformulation of the problem based on a duality argument from
\cite{wu2014jcp}. Convex-concave programs cannot be solved by standard
nonlinear programming approaches which aim to minimize some objective
subject to constrains. They can be treated as finite dimensional
monotone variational inequalities and they can be solved using the
primal-dual interior-point outlined in \cite{ralph2000}.

The reversible MLE problem is a nonlinear programming problem with a
convex objective and non-convex constraints. The number of unknowns in
the problem is quadratic in the number of states of the chain. The
dual problem has only linear constraints and the number of unknowns
grows linearly with the number of states of the chain. The
reformulation can also be applied in order to solve a number of
related MLE problems arising if additional information about the chain
is available \emph{a priori}. A broader class of interesting MLE
problems for reversible Markov chains can thus be solved.

In \cite{wu2014jcp, wu2014siam} the reversible MLE problem has been
extended to the discrete transition matrix reweighting analysis method
(dTRAM). For dTRAM, simulation data at multiple biasing conditions,
also called thermodynamic states, is collected in order to efficiently
estimate the stationary vector at the unbiased condition. A positive
reweighting transformation relates each stationary vector at a biased
condition to the stationary vector at the unbiased condition. This
coupling between unbiased and biased condition makes it possible to
combine the information from all ensembles into the desired estimate
for the unbiased situation.

The dTRAM problem was previously solved through an application of a
self consistent iteration procedure to the dual reformulation
\cite{wu2014jcp}. This approach can require a large number of
iterations to converge. We show that the convex-concave reformulation
of the reversible MLE problem can be extended to also cover the dTRAM
problem. The resulting convex-concave program can be solved using the
algorithm outlined in \cite{ralph2000}. The large linear systems
arising during the computation of the search direction can be
efficiently solved using a Schur complement approach similar to the
one outlined in \cite{zavala2008, kang2014}. The resulting algorithm
achieves a significant speed-up compared to the self consistent
iteration.

%% file: theory.tex
\section{Markov chain estimation}
A Markov chain on a finite state space is completely characterized by
a square matrix of conditional probabilities,
$P = (p_{ij}) \in \mathbb{R}^{n \times n}$. The entry $p_{ij}$ is the
probability for the chain to make a transition to state $j$ given that
it currently resides in state $i$. The matrix $P$ is stochastic,
i.e. $\sum_{j} p_{ij} = 1$ for all $i$. If $P$ is irreducible then
there exists a unique vector, $\pi=(\pi_i) \in \mathbb{R}^n$, of positive probabilities
such that $\pi$ is invariant under the action of $P$,
$\pi^T P = \pi^T$. The vector $\pi$ is called the stationary vector of
the chain.

If there is a vector, $\pi$, of probabilities for which $P$
fulfills the following detailed balance condition,
\begin{equation}
  \label{eq:detailed_balance}
  \pi_i p_{ij} = \pi_j p_{ji}
\end{equation}
then the chain is a reversible Markov chain with stationary vector
$\pi$, \cite{levin2009}.

In Markov chain estimation one is interested in finding an optimal
transition matrix estimate $P$ from a given finite observation
$X=\{X_0,X_1,\dotsc,X_N\}$ of a Markov chain with unknown transition
matrix. The matrix of transition counts $C=(c_{ij})$ together with
the initial state $X_0=x_0$ is a minimal sufficient statistics for
the transition matrix \cite{denny1978}. The element $c_{ij}$ denotes
the observed number of transitions between state $i$ and state $j$ in
$X$.  The matrix $P$ is optimal if it maximizes the following
log-likelihood
\begin{equation}
  \label{eq:logl}
  L(C|P) = \sum_{i,j} c_{ij} \log p_{ij}.
\end{equation}
For finite ensembles consisting of finite length observations one can
simply add the matrices of transition counts for each observation. The
accumulated counts together with the empirical measure of the initial
states is then a sufficient statistics for the finite ensemble of
observations.

For reversible Markov chain estimation one constrains the general
Markov chain MLE problem to the set of all stochastic matrices for
which detailed balance with respect to some vector of probabilities
holds. Thus we can find the reversible MLE transition matrix from the
following nonlinear program,
\begin{equation}
  \label{eq:mle_rev_primal}
  \begin{aligned}
    & \underset{\pi, P}{\min} & & -\sum_{i, j} c_{ij} \log p_{ij} \\
    & \text{subject to} & & p_{ij} \geq 0, \enspace \sum_{j} p_{ij}=1,
    \enspace \pi_i > 0, \enspace \sum_{i} \pi_i = 1, \enspace \pi_i p_{ij} = \pi_j p_{ji}.  \\
  \end{aligned}
\end{equation}

In \cite{wu2014jcp, wu2014siam} problem \eqref{eq:mle_rev_primal} has
been extended to the discrete transition matrix reweighting analysis
method (dTRAM). For dTRAM, simulation data at multiple thermodynamic
states $\alpha=0,\dots,M$ is collected in order to efficiently
estimate the stationary vector at the unbiased condition,
$\alpha=0$. A positive reweighting transformation relates the
stationary vector at the biased condition, $\alpha>0$, to the
stationary vector at the unbiased condition,
\begin{equation}
  \label{eq:reweighting_transformation}
  \pi^{(\alpha)}_i = U^{(\alpha)}_i \pi_i^{(0)} = \exp(u^{(\alpha)}_i) \pi_i^{(0)}.
\end{equation}
This coupling allows us to combine the information from all ensembles
into the estimate for $\pi^{(0)}$. 

The dTRAM problem consists of reversible MLE problems for each
thermodynamic state coupled via the reweighting transformation
\eqref{eq:reweighting_transformation}. The desired stationary vector
can be obtained as the optimal point of the following nonlinear
program,
\begin{equation}
  \label{eq:dTRAM_primal}
  \begin{aligned}
    & \min_{\pi^{(\alpha)}, P^{(\alpha)}} & & -\sum_{\alpha} \sum_{i, j} c_{ij}^{(\alpha)} \log p_{ij}^{(\alpha)} \\
    & \text{subject to} & & p_{ij}^{(\alpha)} \geq 0, \enspace \sum_{j} p_{ij}^{(\alpha)} = 1,
    \enspace \pi^{(\alpha)}_i > 0, \enspace \sum_{i} \pi^{(\alpha)}_i = 1, \\
    & & & \pi^{(\alpha)}_i p^{(\alpha)}_{ij} = \pi^{(\alpha)}_{j}p^{(\alpha)}_{ji}, 
    \enspace \pi_i^{(\alpha)} = U_i^{(\alpha)} \pi_i^{(0)}.
  \end{aligned}
\end{equation}


We show that the convex-concave reformulation of the reversible MLE
problem can be extended to derive an efficient numerical algorithm for
the solution of the dTRAM problem. Additional structure in the linear
systems arising during the primal-dual iteration can be used so that
the problem can be solved efficiently for many coupled chains.


\section{Dual of the reversible MLE problem}
In \cite{wu2014jcp} a duality argument was used to show that finding
the MLE of \eqref{eq:mle_rev_primal} for given positive weights
$\pi_i$ is equivalent to the following concave maximization problem,
\begin{equation}
  \label{eq:mle_revpi_jcp}
  \begin{aligned}
    & \underset{x}{\max} & & \sum_{i,j} c_{ij} \log(\pi_i x_j + \pi_j x_i) - \sum_{i,j} c_{ij}\log \pi_j - \sum_{i} x_i \\
    & \text{subject to} & & x_i \geq 0.
  \end{aligned}
\end{equation}
The $x_i$ correspond to the Lagrange multipliers for the row
normalization constraint in the primal problem
\eqref{eq:mle_rev_primal}. The optimal transition probabilities can be
recovered according to
\begin{equation}
  \label{eq:pij_optimal_jcp}
  p^{*}_{ij} = \frac{(c_{ij}+c_{ji})\pi_j}{\pi_i x^{*}_j + \pi_j x^{*}_i}, \quad j \neq i.
\end{equation}
The vector $x^{*}$ denotes the optimal point of
\eqref{eq:mle_revpi_jcp} and the diagonal entries $p_{ii}^{*}$ are determined by the row
normalization condition. It is clear that $p_{ij}^{*}$ is a proper
probability irrespective of the normalization of the weights since any
scaling of $\pi_i$ cancels out in \eqref{eq:pij_optimal_jcp}.

In \cite{wu2014jcp} the inequality constraints on $x_i$ were not made
explicit. The non-negativity requirement can be seen from the
following splitting of the Lagrangian $L_{\pi}$ in \cite{wu2014jcp},
\begin{equation}
  \label{eq:lagrangian_mle_revpi_jcp}
  \begin{aligned}
    L_{\pi}(P, \lambda, \nu) = & - \sum_{i,j \in I} c_{ij} \log p_{ij} + \sum_{i,j \in I} (\pi_i (\lambda_{ij} - \lambda_{ji}) + x_i) p_{ij}  \\
    & + \sum_{i,j \notin I} (\pi_i (\lambda_{ij} - \lambda_{ji}) + x_i) p_{ij} -\sum_{i} x_i
  \end{aligned}
\end{equation}
with index set $I = \{(i,j)|c_{ij}>0\}$ and the
constraint $p_{ij}\geq 0$. The value $\min_{x} L_{\pi}$ is
not bounded from below if
$\pi_i (\lambda_{ij} - \lambda_{ji}) + x_i<0$ for some
$(i,j) \notin I$. Therefore $x_i \geq 0$ for all $(i, i) \notin I$.
It is also not bounded from below if
$\pi_i (\lambda_{ij} - \lambda_{ji}) + x_i \leq 0$ for some
$(i,j) \in I$, so that $x_i >0 $ for all $(i, i) \in I$,

Using the dual function from \cite{wu2014jcp} the reformulation of the
reversible MLE problem, \eqref{eq:mle_rev_primal}, as a saddle-point
problem with constraints is 
\begin{equation}
  \label{eq:mle_rev_jcp}
  \begin{aligned}
  & \underset{\pi}{\min} \; \underset{x}{\max} & & \sum_{i,j} c_{ij} \log(\pi_i x_j + \pi_j x_i) - \sum_{i,j} c_{ij}\log \pi_j - \sum_{i} x_i \\
  & \text{subject to} & & x_i \geq 0, \enspace \pi_i > 0, \enspace \sum_{i} \pi_i = 1.
  \end{aligned}
\end{equation}
\label{eq:mle_rev_jcp} is concave in $x$ but non-convex in $\pi$. The
problem can however be easily cast into a convex-concave form by the
following change of variables,
\begin{equation}
  \label{eq:probabilities_and_energies}
  \pi_i \propto e^{y_i},
\end{equation}
and by replacing the normalization condition with the simpler constraint
\begin{equation}
  \label{eq:uniqueness_constraint}
  y_1 = 0.
\end{equation}
The constraint in \eqref{eq:uniqueness_constraint} removes the
invariance of the objective in \eqref{eq:mle_rev_jcp} with respect to
a constant shift of $y$. Proper stationary probabilities $\pi_i$ can
be obtained from the new variables $y_i$ according to
\eqref{eq:probabilities_and_energies} followed by straightforward
normalization. The variable $y_i$ is the negative free energy of the
state $i$.

The final form of the dual reversible MLE problem is 
\begin{equation}
  \label{eq:mle_rev_dual}
  \begin{aligned}
    & \underset{y}{\max} \; \underset{x}{\min} & &
     -\sum_{i, j} c_{ij} \log \left(x_i e^{y_j}+ x_j e^{y_i}\right) + \sum_i x_i + \sum_{i,j} c_{ij} y_j \\
    & \text{subject to} & & x_i \geq 0, \enspace  y_1 = 0. 
  \end{aligned}
\end{equation}
The objective in \eqref{eq:mle_rev_dual} is convex in $x$ and concave
in $y$. The feasible set is convex so that \eqref{eq:mle_rev_dual} is
a convex-concave program.

For a given state space with $n$ states the original reversible MLE
problem \eqref{eq:mle_rev_primal}, a non-convex constrained
minimization problem in $\mathcal{O}(n^2)$ unknowns, is reduced to a
convex-concave programming problem in $\mathcal{O}(n)$ unknowns with
simple constraints. 


\subsection{Scaling}
We observe that the number of iterations needed for the solution of
\eqref{eq:mle_rev_dual} using the algorithm from \cite{ralph2000} can
be drastically reduced by scaling the count-matrix by a constant
factor $\gamma$ chosen as
\begin{equation}
  \label{eq:scaling}
  \gamma = \left( \underset{i, j}{\max} \, c_{ij} \right)^{-1}.
\end{equation}

With scaled entries $\tilde{c}_{ij}=\gamma c_{ij}$ and scaled variables
$\tilde{x}=\gamma x$, $\tilde{y} = y$ we have
\begin{equation}
  \label{eq:objective_scaled}
  \tilde{f}_0(\tilde{x}, \tilde{y}) = \gamma f_0(x, y) + \text{const.}
\end{equation}

The constraints in \eqref{eq:mle_rev_dual} are invariant under the
scaling so that the optimal point for \eqref{eq:mle_rev_dual} can be
obtained from the optimal solution to the scaled problem.

The resulting stationary probabilities as well as the transition
probabilities are invariant under the scaling,
\begin{equation}
  \label{eq:scaling6}
  \tilde{p}_{ij} = \frac{(\tilde{c}_{ij} + \tilde{c}_{ji})e^{\tilde{y}_j}}{\tilde{x}_i e^{\tilde{y}_j} + \tilde{x}_j e^{\tilde{y}_i}} 
  = \frac{({c}_{ij} + {c}_{ji})e^{{y}_j}}{{x}_i e^{{y}_j} + {x}_j e^{{y}_i}} = p_{ij}.
\end{equation}

\subsection{Special cases and extensions}
The reversible estimation problem with fixed stationary vector $\pi$
\begin{equation}
  \label{eq:mle_revpi_primal}
  \begin{aligned}
    & \underset{P}{\min} & & -\sum_{i, j} c_{ij} \log p_{ij} \\
    & \text{subject to} & & p_{ij} \geq 0, \enspace \sum_{j} p_{ij}=1, \enspace \pi_i p_{ij} = \pi_j p_{ji}   \\
  \end{aligned}
\end{equation}
is a convex problem and can efficiently be solved in its dual
formulation \eqref{eq:mle_revpi_jcp} using an interior-point method
for convex programming problems.

The reversible estimation problem with partial information about the
stationary vector
\begin{equation}
  \label{eq:mle_revpieq_primal}
  \begin{aligned}
    & \underset{\pi, P}{\min} & & -\sum_{i, j} c_{ij} \log p_{ij} \\
    & \text{subject to} & & p_{ij} \geq 0, \enspace \sum_{j} p_{ij} = 1, 
    \enspace \pi_i > 0, \enspace \sum_{i} \pi_i = 1, \\
    & & & \pi_i p_{ij} = \pi_{j}p_{ji}, \enspace \pi_{i} = \nu_i \enspace i \in I,
  \end{aligned}
\end{equation}
with $I \subsetneq \{1,\dots,n\}$ and given positive weights
$(\nu_i)_{i \in I}$ can be solved via its dual
\begin{equation}
  \label{eq:mle_revpieq_dual}
    \begin{aligned}
      & \underset{y}{\max} \; \underset{x}{\min} & & 
      -\sum_{i, j} c_{ij} \log \left(x_i e^{y_j}+ x_j e^{y_i}\right) + \sum_i x_i + \sum_{i,j} c_{ij} y_j \\
    & \text{subject to} & & x_i \geq 0, \enspace y_i = \log \nu_i \enspace i \in I.
  \end{aligned}
\end{equation}

The reversible estimation problem with bound-constrained information
about the stationary vector
\begin{equation}
  \label{eq:mle_revpiineq_primal}
  \begin{aligned}
    & \underset{\pi, P}{\min} & & -\sum_{i, j} c_{ij} \log p_{ij} \\
    & \text{subject to} & & p_{ij} \geq 0, \enspace \sum_{j} p_{ij} = 1,
    \enspace \pi_i > 0, \enspace \sum_{i} \pi_i = 1, \\
    & & & \pi_i p_{ij} = \pi_{j}p_{ji}, \enspace \eta_i \leq \pi_{i} \leq \xi_i \enspace i \in I. 
  \end{aligned}
\end{equation}
with $I \subseteq \{1,\dots,n\}$ and given positive bounds
$(\eta_i)_{i \in I}$, $(\xi_i)_{i \in I}$ can be solved via the dual
\begin{equation}
  \label{eq:mle_revpiineq_dual}
    \begin{aligned}
      & \underset{y}{\max} \; \underset{x}{\min} & &
    -\sum_{i, j} c_{ij} \log \left(x_i e^{y_j}+ x_j e^{y_i}\right) + \sum_i x_i + \sum_{i,j} c_{ij} y_j \\
     & \text{subject to} & & x_i \geq 0, \enspace \log \eta_i \leq y_i \leq \log \xi_i \enspace i \in I. \\
  \end{aligned}
\end{equation}

The two problems \eqref{eq:mle_revpieq_dual},
\eqref{eq:mle_revpiineq_dual} are convex-concave programming
problems. Nonlinear, convex inequality and linear equality constraints
possibly coupling $x$ and $y$ can also be treated within the
algorithmic framework of \cite{ralph2000}. A special case with
possible interest for applications are bound constraints on the
integrated stationary weights on subsets $S \subseteq \{1,\dots,n\}$,
\begin{equation}
  \label{eq:setprob_bound}
  \sum_{i \in S} \pi_i \leq \nu.
\end{equation}
Equation \eqref{eq:setprob_bound} can be expressed in terms of variables $y_i$ as
\begin{equation}
  \label{eq:setlogprob_bound}
  \log \sum_{i \in S} e^{y_i} \leq \log \nu_k, 
\end{equation}
The logarithm of a sum of exponentials is a convex function, \cite{boyd2004}.

\subsection{dTRAM}
We can apply the duality argument to each thermodynamic state in
\eqref{eq:dTRAM_primal} and introduce the coupling between different
ensembles, \eqref{eq:reweighting_transformation}, through linear
equality constraints. The resulting convex-concave programming problem is
\begin{equation}
  \label{eq:dTRAM_dual}
  \begin{aligned}
    & \underset{y^{(\alpha)}}{\max} \quad \underset{x^{(\alpha)}}{\min} & & 
    -\sum_{\alpha} \sum_{i, j} c_{ij} \log \left(x^{(\alpha)}_i e^{y^{(\alpha)}_j}+
      x_j^{(\alpha)} e^{y^{(\alpha)}_i}\right) + \sum_i x_i^{(\alpha)} + \sum_{i,j} c_{ij} y^{(\alpha)}_j \\
    & \text{subject to} & & x_i^{(\alpha)} \geq 0, \enspace y_i^{(\alpha)} - y_i^{(0)} = u^{(\alpha)}_i,
    \enspace y_1^{(0)} = 0.
  \end{aligned}
\end{equation}

The number of iterations required to solve the dTRAM problem is also
greatly reduced by scaling each count-matrix according to
\begin{equation}
  \label{eq:scaling_dtram1}
  \tilde{c}^{(\alpha)}_{ij} = \gamma c_{ij}^{(\alpha)}
\end{equation}
with 
\begin{equation}
  \label{eq:scaling_dtram2}
  \gamma = \underset{\alpha, i, j}{\max} \; c^{(\alpha)}_{ij}
\end{equation}

As for the reversible MLE problem a larger class of related dTRAM
problems can be solved by augmenting the dual problem
\eqref{eq:dTRAM_dual} with convex constraints, e.g. dTRAM with partial
or bound constrained information about the unbiased stationary
vector. It must be ensured that the additional constraints on the
biased stationary probabilities do not result in an infeasible
problem, i.e. the reweighting condition
\eqref{eq:reweighting_transformation} and the constraints cannot be
fulfilled simultaneously.

\section{Convex-concave programs and variational inequalities}
A convex-concave program is the following saddle point problem,
\begin{equation}
  \label{eq:convex_concave_program} 
  \begin{aligned}
  & \max_y \min_x  & & f(x, y) \\
  & \text{subject to} & & (x, y) \in \mathcal{K}
  \end{aligned}
\end{equation}
with $f$ convex in $x$, concave in $y$, and
$\mathcal{K} \subseteq \mathbb{R}^n$ a convex set.

Convex-concave programs can be treated as special cases of
finite-dimensional variational inequality (VI) problems,
\cite{facchinei2007}: For a given feasible set
$\mathcal{K} \subseteq \mathbb{R}^{n}$ and a mapping
$\Phi: \mathcal{K} \to \mathbb{R}^{n}$ find a point
$z^{*} \in \mathcal{K}$ such that
\begin{equation}
  \label{eq:VI_problem}
  (z-z^{*})^{T}\Phi(z^{*}) \geq 0 \quad \forall z \in \mathcal{K}.
\end{equation}
Any point $z^{*}$ satisfying \eqref{eq:VI_problem} is a solution or
optimal point for the VI. The convex-concave program is
cast into the VI-form by defining
\begin{equation}
  \label{eq:VI_for_cvx_cov}
    \Phi(z) = \left(\begin{array}{c} \nabla_x f(x, y) \\ -\nabla_y f(x, y) \end{array}\right), \quad z=(x, y).
\end{equation}

A mapping $\Phi$ is said to be monotone if 
\begin{equation}
  \label{eq:monotone_mapping}
  (z'-z)^T (\Phi(z') - \Phi(z)) \geq 0 \quad \forall z',z \in \mathcal{K}.
\end{equation}
Monotonicity of \eqref{eq:VI_for_cvx_cov} follows from the
convex-concave property of $f$.

If $\mathcal{K}$ is a convex polyhedral set, i.e. solely defined in terms of
linear equalities and inequalities,
\begin{equation}
  \label{eq:polyhedral_set}
  \mathcal{K} = \{z \in \mathbb{R}^{n}|Az-b=0, \; Gz-h \leq 0\},
\end{equation}
then $z$ solves the VI \eqref{eq:VI_problem} if and only if there are
vectors $\lambda$, $\nu$, $s$, such that the following KKT-conditions
are fulfilled \cite{facchinei2007},
\begin{equation}
  \label{eq:KKT_conditions}
  \begin{aligned}
    \Phi(z)+ A^{T}\nu + G^{T}\lambda &= 0 \\
    Az-b &= 0 \\
    Gz-h + s &= 0 \\
    \lambda^T  s &= 0 \\
    \lambda, s &\geq 0
  \end{aligned}
\end{equation}
The vectors $\lambda$ and $\nu$ are dual variables associated with the
inequality and equality constraints. The vector of slack variables,
$s = (h-Gz)$, transforms the linear inequality constraints for $z$
into simple non-negativity constraints for $s$. Optimality
conditions for convex $\mathcal{K}$ in standard form, i.e. defined by
a finite number of linear equalities and convex inequalities, are also
available, cf. \cite{facchinei2007}.

A direct application of a Newton type method to
\eqref{eq:KKT_conditions} ensuring positivity of $\lambda$ and $s$ is
usually unsuccessful since the solution progress rapidly stagnates
once the iterates approach the boundary of the feasible set. 

A possible strategy to circumvent this problem is numerical
path-following. Instead of attempting a direct solution of
\eqref{eq:KKT_conditions} path-following proceeds by solving a
sequence of problems with perturbed complementarity condition,
\begin{equation}
  \label{eq:KKT_conditions_perturbed}
  \begin{aligned}
    \Phi(z)+ A^{T}\nu + G^{T}\lambda =0 \\
    Az-b = 0 \\
    Gz-h + s = 0 \\
    \lambda^T  s = \mu\\
    \lambda, s \geq 0
  \end{aligned}
\end{equation}
tracing the central path of solutions $z^{*}(\mu)$ towards $z^{*}(0)$
with $\mu \rightarrow 0^{+}$. Perturbing the complementarity condition
ensures that the boundary of the feasible set is not reached
prematurely and the iteration makes good progress along the computed
search direction.

Interior-point methods ensure the positivity of $\lambda$ and $s$
at each step of the iteration. If in addition a strictly feasible
starting point $Az^{(0)}-b = 0$, $Gz^{(0)}-h+s^{(0)}=0$ is used then
all iterates produced by the algorithm lie in the interior of the
feasible region. 

Progress towards a solution of the perturbed KKT-conditions
\eqref{eq:KKT_conditions_perturbed} is usually made by taking steps along
the Newton direction computed from the following linear system,
\begin{equation}
  \label{eq:DKKT_system}
  \left(
    \begin{array}{cccc} 
      D\Phi(z) & A^{T} & G^{T} & 0 \\
      A     & 0    &  0   & 0 \\
      G     & 0    &  0   & I \\
      0     & 0    &  S   & \Lambda
    \end{array}
  \right) \left(
    \begin{array}{c} 
      \Delta z \\ 
      \Delta \nu \\ 
      \Delta \lambda \\
      \Delta s 
    \end{array}
  \right) = -\left(
    \begin{array}{c}
      \Phi(z) + A^{T}\nu + G^{T}\lambda \\
      Az-b \\
      Gz-h+s \\
      S\Lambda \mathbf{e} -\mu \mathbf{e}
    \end{array}
  \right),
\end{equation}
with diagonal matrices $S=\text{diag}(s_1,s_2,\dotsc)$,
$\Lambda=\text{diag}(\lambda_1,\lambda_2,\dotsc)$, the vector
$\mathbf{e}=(1,1,\dotsc)$, and the perturbation parameter $\mu>0$.

We use the following short-hand notation for the dual
residuum,
\begin{equation}
  \label{eq:res_dual}
  r_{d} = \Phi(z) + A^{T}\nu + G^{T}\lambda,
\end{equation}
the primal residuals,
\begin{equation}
  \label{eq:res_primal}
  \begin{aligned}
    r_{p,1} &= Az-b, \\
    r_{p,2} &= Gz-h+s,
  \end{aligned}
\end{equation}
and the perturbed complementary slackness,
\begin{equation}
  \label{eq:slackness}
  r_c(\mu) = S \Lambda \mathbf{e} - \mu \mathbf{e}.
\end{equation}

Solving the linear system \eqref{eq:DKKT_system} is the most
expensive part of the algorithm. The sparse block structure of
\eqref{eq:DKKT_system} can be used to significantly speed up the
solution process. Elimination of $\Delta s$ and $\Delta \lambda$
reduces \eqref{eq:DKKT_system} to the \emph{augmented system}
\begin{equation}
  \label{eq:augmented_system}
    \left(
      \begin{array}{cc}
        H & A^{T} \\
        A & 0
      \end{array}
    \right)
    \left(
      \begin{array}{c}
        \Delta z \\
        \Delta \nu
      \end{array}
    \right)
   = -
  \left(
    \begin{array}{c}
      r_{d} + G^{T} \Sigma r_{p,2} - G^T S^{-1} r_{c}(\mu) \\
      r_{p,1}
    \end{array}
  \right),
\end{equation}
with diagonal matrix $\Sigma = S^{-1} \Lambda$ and augmented Jacobian
$H = D\Phi + G^{T}\Sigma G$.  The increments $\Delta \lambda$ and
$\Delta s$ can be computed from $\Delta z$,
\begin{equation}
  \begin{aligned}
    \Delta s & = -r_{p, 2} - G \Delta z \\
    \Delta \lambda & = - \Sigma \Delta s - S^{-1} r_{c}(\mu).
  \end{aligned}
\end{equation}

For nonsingular $H$ further elimination of $\Delta z$ from \eqref{eq:augmented_system} is possible. The resulting \emph{normal
  equations} for $\Delta \nu$ are,
\begin{equation}
  \label{eq:normal_system}
    S \Delta \nu = r_2 - AH^{-1} r_1.
\end{equation}
The vectors $r_i$ are the two components of the RHS of
\eqref{eq:augmented_system} and the matrix
$S = \left(A H^{-1} A^{T}\right)$ is the Schur complement of $H$. The
increment $\Delta z$ can then be computed according to
\begin{equation}
  \Delta z = -H^{-1}(r_1 + A^{T} \Delta \nu).
\end{equation}

A singular matrix $H$ can for example occur for an equality-constrained
convex programming problem for which the objective is not strictly
convex. Even if the constraints ensure that the problem has a unique
solution, $H$ will be singular so that the normal equations can not be
formed.

For convex programming problems a non-singular $H$ can be efficiently
factorized using a symmetric positive-definite Cholesky
factorization. In the convex-concave case the Jacobian of the mapping
$\Phi$ is not symmetric,
\begin{equation}
  \label{eq:Jacobian_phi}
  D\Phi(z) = \left(
    \begin{array}{cc}
      \nabla_x \nabla_x f(x, y) & \nabla_y \nabla_x f(x, y)^{T} \\
      -\nabla_y \nabla_x f(x, y) & -\nabla_y \nabla_y f(x, y)
      \end{array}
    \right).
\end{equation}
In that case the augmented system is not symmetric and the Cholesky
factorization can not be used.

A further speed-up in the computation of the Newton direction can be
achieved through the exploitation of sparse or block-sparse structure
possibly present in $D\Phi$, $G$, $A$. In this situation solution via
an iterative method can be particularly efficient if a good
preconditioner is available.

\section{Implementation details}
In order to apply the algorithm in \cite{ralph2000} to the reversible
MLE problem \eqref{eq:mle_rev_dual} we transform the convex-concave
program into the VI form using the mapping
$\Phi=(\nabla_{x}f, -\nabla_y f)$ in \eqref{eq:VI_for_cvx_cov}. The
gradient of the objective in \eqref{eq:mle_rev_dual} is given by
\begin{equation}
  \label{eq:gradient}
  \begin{aligned}
    \partial_{x_k} f &= - \sum_{j}  \frac{(c_{kj} + c_{jk}) e^{y_j}}{x_k e^{y_j} + x_j e^{y_k}} + 1 \\
    \partial_{y_k} f &= - \sum_{j}  \frac{(c_{kj} + c_{jk}) x_j e^{y_k}}{x_k e^{y_j} + x_j e^{y_k}} + \sum_{i} c_{ik}.
  \end{aligned}
\end{equation}

For the computation of the Newton direction we also need the Jacobian
$D\Phi$. The diagonal blocks are given by
\begin{equation}
  \label{eq:Hessian1}
  \begin{aligned}
    \partial_{x_k} \partial_{x_l} f &= \sum_{j} \frac{(c_{kj} + c_{jk}) e^{y_j} e^{y_j}}{(x_k e^{y_j} + x_j e^{y_k})^2}\delta_{k,l} +
    \frac{(c_{kl}+c_{lk}) e^{y_k} e^{y_l}}{(x_k e^{y_l} + x_l e^{y_k})^2}, \\
    \partial_{y_k} \partial_{y_l} f & = -\sum_{j} \frac{(c_{kj} + c_{jk}) x_k e^{y_j} x_j e^{y_k}}{(x_k e^{y_j} + x_j e^{y_k})^2} \delta_{k,l} +
    \frac{(c_{kl}+c_{lk}) x_k e^{y_l} x_l e^{y_k}}{(x_k e^{y_l} + x_l e^{y_k})^2},
  \end{aligned}
\end{equation}
and off-diagonal blocks are given by
\begin{equation}
  \label{eq:Hessian2}
  \begin{aligned}
   \partial_{y_k} \partial{x_l} f &= \sum_{j} \frac{(c_{kj}+c_{jk})e^{y_k} x_j e^{y_j}}{(x_k e^{y_j} + x_j e^{y_k})^2}\delta_{k, l} - 
   \frac{(c_{kl} + c_{lk}) x_k e^{y_k} e^{y_l}}{(x_k e^{y_l} + x_l e^{y_k})^2}, \\
   \partial_{x_k} \partial_{y_l} f &= \partial_{y_l} \partial{x_k} f.
  \end{aligned}
\end{equation}

It is straightforward to encode the equality and inequality
constraints in \eqref{eq:mle_rev_dual} into matrices $A$, $G$ and
vectors $b$, $h$.
\begin{equation}
  \label{eq:mle_rev_A}
  A = (\underbrace{0,\dots,0}_{n},\underbrace{1,0,\dots,0}_{n}),
\end{equation}

\begin{equation}
  \label{eq:mle_rev_b}
  b = 0,
\end{equation}

\begin{equation}
  \label{eq:mle_rev_G}
  G = (-I_{n}, 0_{n}),
\end{equation}

\begin{equation}
  \label{eq:mle_rev_h}
  h = (0,\dots,0)^T
\end{equation}
with $I_{n}$ the identity and $0_{n}$ the zero matrix in $\mathbb{R}^{n\times n}$.

The Jacobian $D\Phi$ is singular because of the invariance
of the objective $f$ under a constant shift of $y$; this is also
true for the augmented Jacobian $H$ since the inequalities act only on
$x$. Therefore the normal equations \eqref{eq:normal_system} cannot
be formed and the search direction has to be computed from the augmented
system \eqref{eq:augmented_system}.

The blocks of $D\Phi$ have the same sparsity pattern as the matrix
$C_s=C+C^T$. This matrix is usually sparse. The augmented Jacobian
differs from the original Jacobian only on the diagonal so that it is
also sparse in a situation in which $C_s$ is sparse. The equality
constraints for the reversible MLE problem do only affect the $y$
variables, i.e. $A = (0, A_y)$. The augmented system,
\eqref{eq:augmented_system}, can be cast into the following symmetric
form,
\begin{equation}
  \label{eq:augmented_system_symmetric}
  \left(
    \begin{array}{ccc}
      H_{xx} & H_{yx} & 0 \\
      H_{yx}^T & -H_{yy} & -A_{y}^{T} \\
      0 & -A_y & 0 
    \end{array}
  \right)    
  \left(
    \begin{array}{c}
      \Delta x \\
      \Delta y \\
      \Delta \nu
    \end{array}    
  \right) = 
  \left(
    \begin{array}{c}
      b_x \\
      -b_y \\
      -b_{\nu}
    \end{array}    
  \right).
\end{equation}
The augmented system matrix, $W$, on the left-hand side of
\eqref{eq:augmented_system_symmetric} is indefinite so that a
symmetric indefinite factorization, \cite{bunch1977}, or the minimum
residual (MINRES) method, \cite{paige1975}, can be used to solve
\eqref{eq:augmented_system_symmetric}. If an iterative method is used,
a suitable preconditioner needs to remove the ill-conditioning due to
the $\Sigma = S^{-1} \Lambda$ term in $H$. MINRES requires a positive
definite preconditioner. We use a positive definite diagonal
preconditioning matrix, $T$, with diagonal entries,
\begin{equation}
  \label{eq:mle_rev_preconditioner}
  t_{ii} = \begin{cases} \lvert w_{ii} \rvert & \mbox{if } \lvert w_{ii} \rvert >0 \\ 1 & \mbox{else } \end{cases}.
\end{equation}

\subsection{dTRAM}
We can also apply the primal-dual interior-point method to the
convex-concave reformulation of the dTRAM problem,
\eqref{eq:dTRAM_dual}. The dTRAM problem consists of a reversible MLE
problem for each thermodynamic state coupled via an equality
constraint. The resulting VI-mapping for dTRAM is given by the vector
\begin{equation*}
  \Phi = (\Phi_0,\dots,\Phi_m).
\end{equation*}
The entry $\Phi_{\alpha}$ is the mapping for the reversible MLE
problem at thermodynamic state $\alpha$. Since $\Phi_{\alpha}$ depends
only on variables $(x^{(\alpha)}, y^{(\alpha)})$ the Jacobian of $\Phi_{\alpha}$ has a block-diagonal structure
\begin{equation*}
  D\Phi = 
  \left(
    \begin{array}{ccc}
      D\Phi_0 &        & \\
                 & \ddots &  \\
                 &        & D\Phi_m \\
    \end{array}
  \right)
\end{equation*}
The matrix $D\Phi_{\alpha}$ is the mapping for the reversible MLE
problem at thermodynamic state $\alpha$. The linear inequality
constraints at different $\alpha$ are decoupled so that $G$ is also
block diagonal,
\begin{equation*}
  G = 
  \left(
    \begin{array}{ccc}
      G_0 &        & \\
                 & \ddots &  \\
                 &        & G_m \\
    \end{array}
  \right).
\end{equation*}
The block $G^{(\alpha)}$ is the matrix of inequality constraints at thermodynamic state $\alpha$,
\begin{equation*}
  G_{\alpha} =  \left(-I_{n}, 0_{n} \right),
\end{equation*}
and $h=0$ is the corresponding RHS. The matrix for the equality
constraints has the following form,
\begin{equation*}
  A = 
  \left(
    \begin{array}{cccc}
      A_0 & 0 & \dots & 0 \\
      A_{1,0} & A_1 & \dots & 0 \\
      \vdots & \vdots & \ddots & \vdots \\
      A_{m,0} & 0 & \dots & A_m
    \end{array}
  \right)
\end{equation*}
with $A_0 = (0,\dots,0,1,\dots,0)$ the constraint matrix for the
unbiased ensemble, $\alpha=0$, and $A_{\alpha} = \left(0_n, I_n \right)$
the constraint matrix at condition $\alpha \neq 0$. The matrix
$A_{\alpha, 0} = \left(0_n, -I_n \right)$ is the coupling matrix between
biased and unbiased ensemble. The corresponding RHS is
\begin{equation*}
  b = 
  \left(
    \begin{array}{c}
      b_0 \\
      \vdots \\
      b_m
    \end{array}
  \right)
\end{equation*}
with $b_0 = 0$, and $b_{\alpha} = (u^{(\alpha)}_i)$ the vector
of energy differences with respect to the unbiased condition.

The block-diagonal form of $D\Phi$ and $G$ can be exploited for the
solution of the augmented system. The block diagonal structure of
$D\Phi$ and $G$ implies a block diagonal structure for $H$,
\begin{equation}
  H = 
  \left(
    \begin{array}{ccc}
      H_1 &  & \\
          & \ddots & \\
          &        & H_m,
    \end{array}
  \right).
\end{equation}
The block
$H_{\alpha} = D\Phi_{\alpha} + G_{\alpha}^T \Sigma_{\alpha}
G_{\alpha}$
is the augmented Jacobian at thermodynamic state $\alpha$.  Using the
block structure of $H$ and $A$, the augmented system
\eqref{eq:augmented_system} can be reordered resulting in the following
linear system,
\begin{equation} \label{eq:augmented_system_dtram}
  \left(
    \begin{array}{cccc}
      W_0   & B_{1, 0}^{T} & \dots & B_{m, 0}^{T} \\
      B_{1, 0} & W_1 & \dots & 0                   \\
      \vdots   & \vdots & \ddots & \vdots             \\
      B_{m, 0} &  0     & \dots  & W_m
    \end{array}
  \right)
  \left(
    \begin{array}{c}
      \Delta \xi_0\\
      \Delta \xi_1 \\
      \vdots \\
      \Delta \xi_m
    \end{array}
  \right) = -
  \left(
    \begin{array}{c}
      \tilde{b}_0 \\
      \tilde{b}_1 \\
      \vdots \\
      \tilde{b}_m     
    \end{array}
  \right)
\end{equation}
The augmented system matrix at condition $\alpha$ is
\begin{equation}
  \label{eq:w_block}
  W_{\alpha} =
  \left(
    \begin{array}{cc}
      H_{\alpha} & A^T_{\alpha} \\
      A_{\alpha} & 0
    \end{array}
  \right).
\end{equation}
The coupling between the biased condition and the unbiased condition is encoded in the matrix
\begin{equation}
  \label{eq:b_block}
  B_{\alpha, 0} = 
  \left(
    \begin{array}{cc}
      0 & 0 \\
      A_{\alpha, 0} & 0
    \end{array}
  \right) \quad \alpha \neq 0.
\end{equation}

The vector
$\Delta \xi_{\alpha} = (\Delta z_{\alpha}, \Delta \nu_{\alpha})$ is
the resulting increment for the augmented system at condition
$\alpha$. The vector $\tilde{b}_{\alpha}$ in
\eqref{eq:augmented_system_dtram} is given by the RHS of the augmented
system at condition $\alpha$,
\begin{equation}
  \tilde{b}_{\alpha} = 
  \left(
    \begin{array}{c}
      r^{(\alpha)}_{d} + G_{\alpha}^T \Sigma_{\alpha} r^{(\alpha)}_{p,2} - G_{\alpha}^T S_{\alpha}^{-1}r^{(\alpha)}_c(\mu) \\
      r^{(\alpha)}_{p,1}
    \end{array}
  \right).
\end{equation}

The arrow-shaped structure of the linear system in
\eqref{eq:augmented_system_dtram} allows us to apply the Schur complement
method, \cite{zavala2008, kang2014}, to eliminate
$\Delta \xi_1,\dots,\Delta \xi_{m}$ and solve the following condensed
system for $\Delta \xi_0$,
\begin{equation}
  \label{eq:condensed_system}
   S \Delta \xi_0 = 
  - \left(\tilde{b}_0 - \sum_{\alpha=1}^{m} B_{\alpha, 0}^{T}W_{\alpha}^{-1}\tilde{b}_{\alpha}\right)
\end{equation}
The Schur complement matrix is
\begin{equation}
  \label{eq:Schur_complement_dTRAM}
  S = \left(W_0 - \sum_{\alpha=1}^{m} B_{\alpha, 0}^{T}W_{\alpha}^{-1}B_{\alpha, 0}\right).
\end{equation}
All other increments can be computed from $\Delta \xi_{0}$ via
\begin{equation}
  \label{eq:Schur_complement_subst}
  \Delta \xi_{\alpha} = -W_{\alpha}^{-1}\left(\tilde{b}_{\alpha} + B_{\alpha, 0}\Delta \xi_0 \right)
\end{equation}

For a system with $n$ states at $m$ thermodynamic conditions the
complexity for a direct factorization of the Newton system
\eqref{eq:DKKT_system} is $\mathcal{O}(m^{3}n^{3})$. The Schur
complement approach reduces complexity to $\mathcal{O}(m n^3)$.  In
addition, assembly of the Schur complement in
\eqref{eq:Schur_complement_dTRAM} and solution of
\eqref{eq:Schur_complement_subst} can be easily paralellized.

As for the reversible MLE case, the blocks of $D\Phi_{\alpha}$ have the
same sparsity pattern as the matrix $C_s^{(\alpha)}=C^{(\alpha)}+C^{(\alpha)T}$. The
same is true for the augmented Jacobian $H_{\alpha}$ except for the
diagonal. Since $C_s^{(\alpha)}$ is usually sparse we use
a sparse LU method to factor the augmented system matrices
$W_{\alpha}$ for $\alpha>0$.  The direct assembly of the Schur complement
in \eqref{eq:Schur_complement_dTRAM} is expensive since the
computation of $W_{\alpha}^{-1}B_{\alpha, 0}$ requires 
$\mathcal{O}(n)$ solves.

If an iterative method is used to solve the condensed system
\eqref{eq:condensed_system} one would like to avoid assembly of the
Schur complement $S$ in \eqref{eq:Schur_complement_dTRAM}
all together. Instead only few matrix-vector products involving $S$ should
be computed. As for the reversible MLE case, we can transform the
condensed system into a symmetric indefinite form and use MINRES to
obtain a solution. Obtaining a good preconditioner without
explicit assembly of $S$ is difficult. We use the probing method outlined in
\cite{chan1992} to obtain an approximation of the diagonal of $S$
using only few matrix-vector products. We then construct a positive
definite diagonal preconditioning matrix $T$ with entries
\begin{equation*}
  t_{ii} = \begin{cases} \lvert \tilde{s}_{ii} \rvert & \mbox{if } \lvert \tilde{s}_{ii} \rvert >0 \\ 1 & \mbox{else } \end{cases}.
\end{equation*}
The entry $\tilde{s}_{ii}$ denotes the diagonal entry estimated by the
probing approach.

The Schur complement based solution can also be applied to the dTRAM
problem with additional constraints whenever those constraints do not
couple different biasing conditions.

%% file: results.tex
\section{Results}
Below we report results for the primal-dual interior-point (Newton-IP)
and the self consistent iteration (SC-iteration) approach to solving
the reversible MLE and dTRAM problem. We compare the efficiency of
both algorithms for a number of examples. Using iterative methods for
the solution of the linear systems arising in the Newton-IP approach
we achieve a similar scaling behavior as for the SC-iteration. We
demonstrate that the Newton-IP approach offers a significant speedup
for nearly all examples.

\subsection{Reversible MLE}
In \autoref{tab:mle_rev} we compare the performance of the algorithm
for different example data-sets. The count matrix was estimated from
the full data set using the sliding-window method
\cite{prinz2011}. The tolerance indicating convergence was
$\text{tol}=10^{-12}$ for both algorithms. Both methods exhibit a
subquadratic scaling in the number of states. The Newton-IP method is
able to achieve a significant speed-up over the SC-iteration for all
examples except for the pentapeptide data.

\begin{figure}
  \centering
  \subfloat[]
  {
    \includegraphics[width=0.5\textwidth]{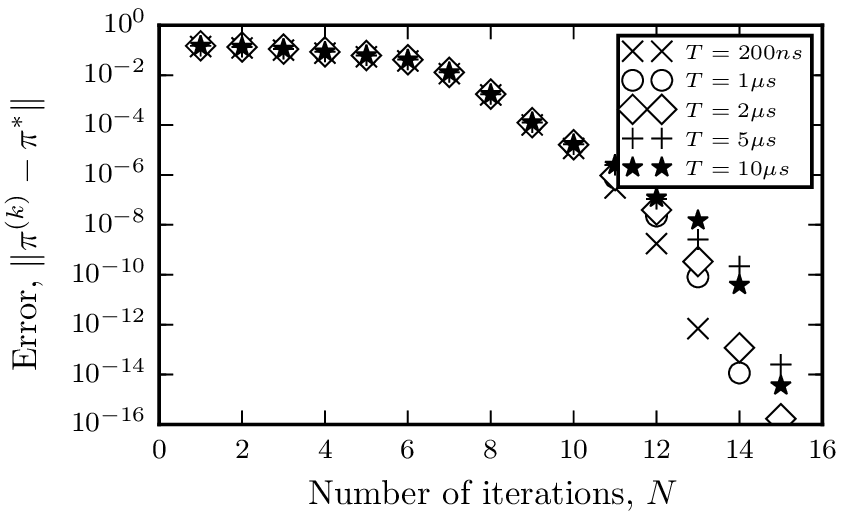}
  }
  \subfloat[]
  {
    \includegraphics[width=0.5\textwidth]{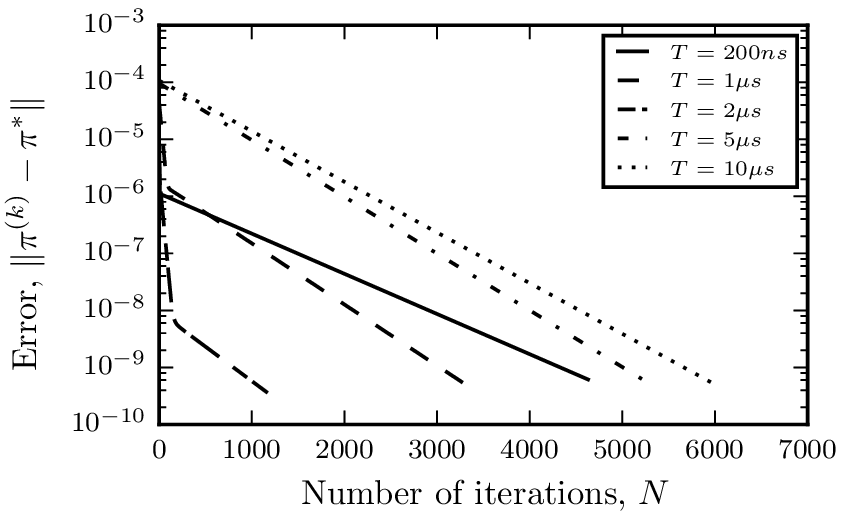}
  }

  \subfloat[]
  {
    \includegraphics[width=0.5\textwidth]{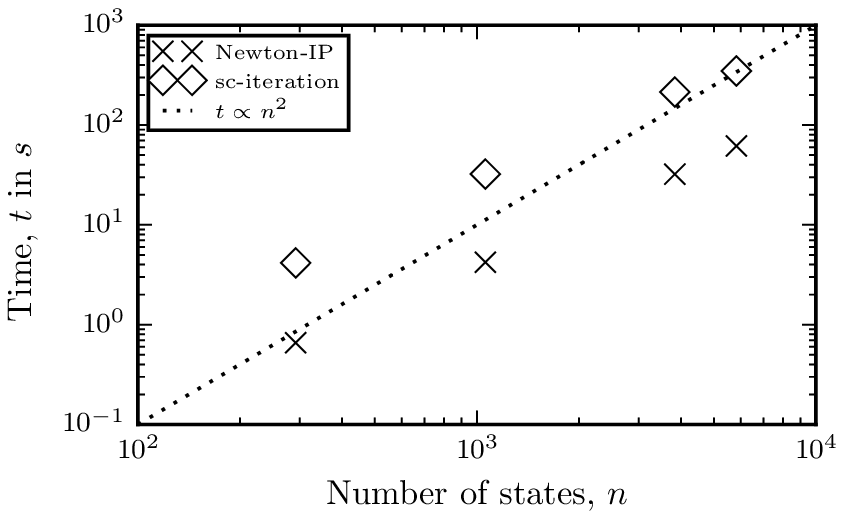}
  }
  
  \caption{Comparison of Newton interior-point method, a), and
    self-consistent iteration, b) for the alanine dipeptide
    example. Convergence is plotted for different data sets
    corresponding to different amounts of total simulation time. The
    vector $\pi^{*}$ is a reference stationary distribution obtained
    from the converged Newton interior-point method. The Newton
    interior-point method converges superlinearly, the
    self-consistent iteration converges linearly. The number of
    required iterations is very sensitive to the input data set for the
    SC-iteration while the Newton-IP method is only mildly
    affected. c) Both methods exhibit a subquadratic scaling in the
    number of states. The Newton-IP method achieves a significant
    speed-up over the SC-iteration.}
  \label{fig:ip_vs_sc_20x20}
\end{figure}

In \autoref{fig:ip_vs_sc_20x20} we show the performance of both
methods for the alanine dipeptide system with 361 states. For the
SC-iteration the number of iterations required to converge to a given
tolerance is very variable across different data sets. The total number
of iterations required to converge deteriorates with increasing amount
of input data. For the Newton-IP method the required number of
iterations is consistent across all data sets. Both methods exhibit
subquadratic scaling in the number of observed states.

\begin{table}
  \caption{Reversible MLE problem. Newton-IP algorithm
vs. SC-iteration. We report the number of states $N$, the growth
factor for states $N/n$ ($n$ is the number of states in the previous
row), the total algorithm run time $T$ (in seconds), the growth factor
for run time $T/t$ ($t$ is the run time in the previous row), the
scaling exponent for run time with increasing number of states $p$, ($T
\propto N^p$), and the speedup of the Newton-IP method over the SC-iteration
SC/IP. The scaling is subquadratic for both methods. The
Newton-IP algorithm achieves a significant speed-up over the
SC-iteration for all examples except the pentapeptide.}
  \centering
  \begin{tabular*}{\textwidth}{rrrrrrrrrr}
    \toprule
    System & $N$ & $N/n$  & \multicolumn{3}{c}{Newton-IP} & \multicolumn{3}{c}{SC-iteration} & SC/IP \\
           &                 &           & $T$ & $T/t$  & $p$ & $T$ & $T/t$ & $p$ & \\    
    \midrule
    \multirow{4}{*}{Three-well} & 361 &  & 1.1 &  &  & 4.6 &  &  & 4.0 \\
           &2134 & 5.9 & 7.3 & 6.4 & 1.0 & 75.1 & 16.2 & 1.6 & 10.2 \\
           &8190 & 3.8 & 56.8 & 7.7 & 1.5 & 400.3 & 5.3 & 1.2 & 7.0 \\
           &29618 & 3.6 & 286.8 & 5.0 & 1.3 & 1076.9 & 2.7 & 0.8 & 3.8 \\
    \midrule
    \multirow{4}{*}{Alanine} & 292 &  & 0.7 &  &  & 4.2 &  &  & 6.3 \\
           & 1059 & 3.6 & 4.2 & 6.4 & 1.4 & 32.3 & 7.8 & 1.6 & 7.6 \\
           & 3835 & 3.6 & 32.2 & 7.6 & 1.6 & 214.0 & 6.6 & 1.5 & 6.6 \\
           & 5826 & 1.5 & 61.8 & 1.9 & 1.6 & 347.7 & 1.6 & 1.2 & 5.6 \\
    \midrule
    \multirow{4}{*}{Pentapeptide} & 250 &  & 0.6 &  &  & 0.2 &  &  & 0.4 \\
           & 500 & 2.0 & 1.2 & 1.9 & 0.9 & 0.6 & 2.4 & 1.3 & 0.5 \\
           & 1000 & 2.0 & 3.6 & 3.0 & 1.6 & 1.0 & 1.8 & 0.9 & 0.3 \\
           & 2000 & 2.0 & 5.4 & 1.5 & 0.6 & 1.3 & 1.3 & 0.4 & 0.2 \\
    \midrule
    \multirow{4}{*}{Birth death} &100 &  & 1.0 &  &  & 10.4 &  &  & 10.6 \\
           &200 & 2.0 & 2.1 & 2.1 & 1.1 & 34.1 & 3.3 & 1.7 & 16.3 \\
           &500 & 2.5 & 5.8 & 2.8 & 1.1 & 185.3 & 5.4 & 1.8 & 31.7 \\
           &1000 & 2.0 & 13.9 & 2.4 & 1.3 & 338.7 & 1.8 & 0.9 & 24.3 \\
    \bottomrule
  \end{tabular*}
  \label{tab:mle_rev}
\end{table}

\subsection{dTRAM }
In \autoref{tab:dtram} we compare the performance of the Newton-IP and
the SC-iteration for different examples.  The count matrix was
estimated from the full data set using the sliding-window method
\cite{prinz2011}. The tolerance indicating convergence was
$\text{tol}=10^{-10}$ for both algorithms. The Newton-IP method is
more efficient for all three examples and achieves a dramatic speed-up
(orders of magnitude). The Schur complement probing approach is
successful for the alanine and the doublewell umbrella sampling
example. For the multi-temperature example the Schur complement was
assembled and the condensed system was solved using a direct
method. For the SC-iteration method the required time to solve the
multi-temperature example was very large so that computations were
only carried out for two examples with a small number of states.

Both methods scale linearly in the number of thermodynamic states.
The Newton-IP method with Schur complement probing scales at most
quadratic in the number of states. If the Schur complement is
assembled and factored by a direct method the scaling is between
quadratic and cubic. The SC-iteration exhibits quadratic scaling in
the number of states. The Newton-IP method achieves orders of
magnitude speed-up compared to the SC-iteration for all examples.

\begin{figure}
  \centering
  \subfloat[]
  {
    \includegraphics[width=0.5\textwidth]{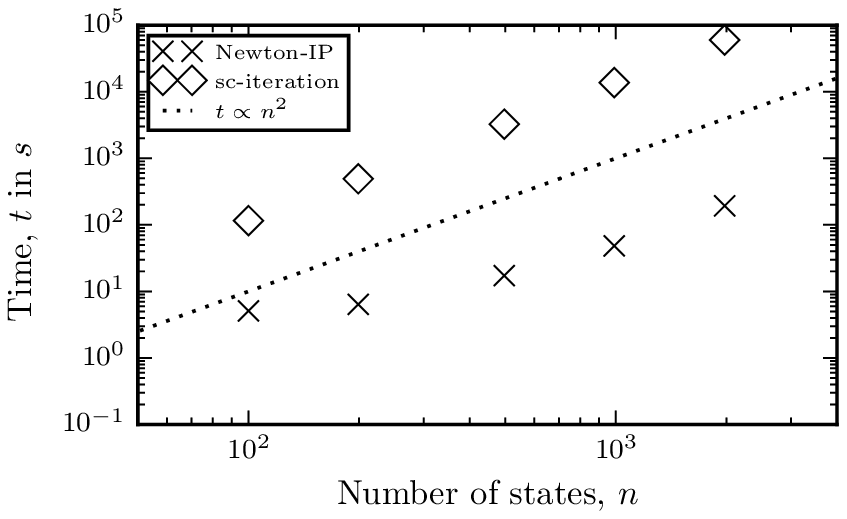}
  }
  \subfloat[]
  {
    \includegraphics[width=0.5\textwidth]{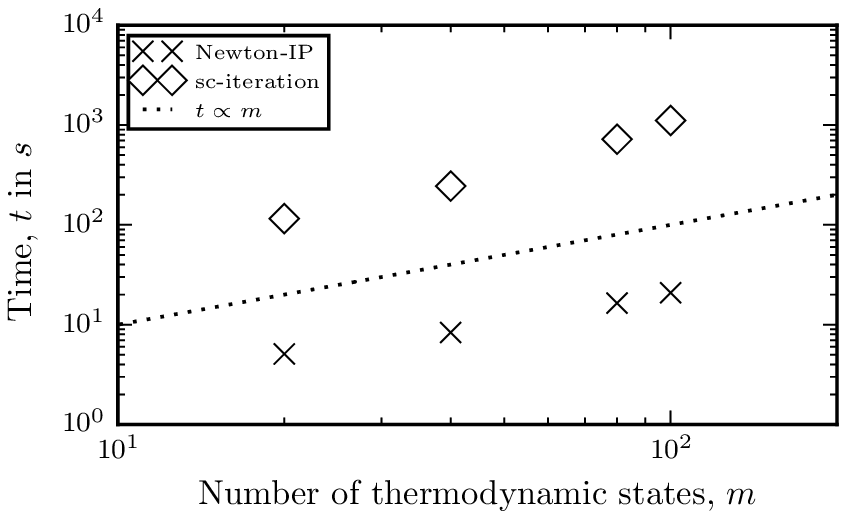}
  }  
  \caption{Comparison of the Newton-IP method and the SC-iteration for
    the dTRAM problem. We show results for the doublewell potential
    with harmonic umbrella forcing. a) Both methods exhibit quadratic
    scaling in the number of states, but the Newton method is up to two
    orders of magnitude faster then the sc iteration. b)
    Scaling is linear in the number of thermodynamic states for both
    methods.}
\label{fig:dtram_ip_vs_sc}
\end{figure}

In \autoref{fig:dtram_ip_vs_sc} we show performance of the Newton-IP
and SC-iteration for the doublewell umbrella-sampling example. The
Newton-IP method achieves a significant speed-up (up to two orders of
magnitude) over the SC-iteration.

\begin{table}
  \centering
  \caption{
    Newton-IP algorithm
    vs. SC-iteration for the dTRAM problem. We report the number of
    states $N$, the number of thermodynamic state $M$, the growth factor
    for states $N/n$ ($n$ is the number of states in the previous row), the
    total algorithm run time $T$ (in seconds), the growth factor for
    run time $T/t$ ($t$ is the run time in the previous row), the scaling
    exponent for run time with increasing number of states $p$, ($T \propto
    N^p$), and the speedup of the Newton IP method over the SC method
    SC/IP. In one case we report instead the growth factor of the number
    of thermodynamic states $M/m$ ($m$ is the number of states in the
    previous row) and the scaling exponent for run time with increasing number
    of thermodynamic states ($T \propto M^p$). Both method scale linearly
    in the number of thermodynamic states. The Newton-IP method with Schur
    complement probing (alanine, doublewell with umbrella sampling) scales
    at most quadratic in the number of states. If the Schur complement is
    assembled and factored by a direct method (doublewell with independent
    temperature sampling) the scaling is between quadratic
    and cubic.  The SC-iteration exhibits quadratic scaling in the number
    of states. The Newton-IP method achieves orders of
    magnitude speed-up compared to the SC-iteration for all examples.
  }
  \begin{tabular*}{\textwidth}{p{1.5cm}rrrrrrrrrr}
    \toprule
    System & $N$ & $M$ & $N/n$  & \multicolumn{3}{c}{Newton-IP} & \multicolumn{3}{c}{SC-iteration} & SC/IP \\
           &     &  &      & $T$ & $T/t$  & $p$ & $T$ & $T/t$ & $p$ & \\ 
    \midrule
    \multirow{2}{\linewidth}{Alanine} & 292 & 40 &   & 34.0 &   &   & 1263.9 &   &   & 37.2 \\
           & 1521 & 40 & 5.2 & 202.4 & 6.0 & 1.1 & 66018.4 & 52.2 & 2.4 & 326.2 \\
    \midrule    
    \multirow{4}{\linewidth}{Doublewell, umbrella} 
           & 100 & 20 &   & 5.1 &   &   & 115.5 &   &   & 22.7 \\
           & 199 & 20 & 2.0 & 6.4 & 1.3 & 0.3 & 492.9 & 4.3 & 2.1 & 77.1 \\
           & 497 & 20 & 2.5 & 17.3 & 2.7 & 1.1 & 3258.4 & 6.6 & 2.1 & 188.7 \\
           & 990 & 20 & 2.0 & 48.3 & 2.8 & 1.5 & 13729.7 & 4.2 & 2.1 & 284.4 \\
           & 1978 & 20 & 2.0 & 193.1 & 4.0 & 2.0 & 59890.5 & 4.4 & 2.1 & 310.1 \\
    \midrule
    \multirow{4}{\linewidth}{Doublewell, umbrella} 
           & 100 & 20 &   & 5.1 &   &   & 115.5 &   &   & 22.7 \\
           & 100 & 40 & 2.0 & 8.3 & 1.6 & 0.7 & 244.5 & 2.1 & 1.1 & 29.3 \\
           & 100 & 80 & 2.0 & 16.5 & 2.0 & 1.0 & 721.1 & 2.9 & 1.6 & 43.8 \\
           & 100 & 100 & 1.2 & 20.9 & 1.3 & 1.1 & 1110.6 & 1.5 & 1.9 & 53.1 \\
    \midrule
    \multirow{4}{\linewidth}{Doublewell, multi-temperature}
           &100 & 16 &   & 3.7 &   &   & 12223.2 &   &   & 3285.8 \\
           &200 & 16 & 2.0 & 10.7 & 2.9 & 1.5 & 50446.2 & 4.1 & 2.0 & 4705.8 \\
           &500 & 16 & 2.5 & 79.8 & 7.4 & 2.2 &   &   &   &   \\
           &1000 & 16 & 2.0 & 544.5 & 6.8 & 2.8 &   &   &   &   \\
    \bottomrule
  \end{tabular*}
  \label{tab:dtram}
\end{table}




%% file: discussion.tex
\section{Conclusion}
We show that the problem of finding the maximum likelihood reversible
transition matrix on a finite state space is equivalent to a
convex-concave programming problem with a much smaller number of
unknowns and constraints.

The primal-dual interior-point method for monotone variational
inequalities outlined in \cite{ralph2000} can be used to efficiently
solve the arising convex-concave program. For a number of examples the
proposed algorithm significantly speeds up the computation of the
reversible MLE compared to a previously proposed fixed-point
iteration.

The convex-concave reformulation makes it possible to efficiently
solve a number of related problems arising in the context of
reversible Markov chain estimation.

One application of special interest is statistical reweighting of data
from multiple ensembles via the dTRAM method \cite{wu2014jcp}. We
extend the convex-concave reformulation to the dTRAM problem so that
it can also be solved by a primal-dual interior-point method. We show
that the arising linear systems can be efficiently solved using a
Schur complement approach. The outlined algorithm is shown to
significantly speed up the solution process compared to a previously
proposed fixed-point iteration.

Similar to the reversible MLE problem a number of related dTRAM
problems can be solved using our method. The efficient linear solution
of the arising Newton systems using the Schur-complement method can be
retained no additional coupling between the different thermodynamic
ensembles is introduced.

The investigation of efficient preconditioning techniques for the
presented problems remains a topic for future research. Obtaining a
good preconditioner for the Schur complement without direct assembly is
of special interest for the dTRAM problem.

%% file: acknowledge.tex
\section*{Acknowledgments}
The authors would like to thank C. Wehmeyer and F. Paul for
stimulating discussions. B. T.-S. thanks E. Pipping and C. Gr\"{a}ser
for valuable comments and suggestions.